\documentclass[epj]{webofc}
\usepackage[varg]{txfonts}   
\woctitle{MESON2016 - the 14$^\textrm{th}$ International Workshop on Meson Production, Properties and Interaction}
\begin{document}
\selectlanguage{english}
\title{The KLOE-2 experiment at DA$\Phi$NE}

\author{M.~Silarski\inst{1}\fnsep\thanks{\email{michal.silarski@lnf.infn.it}}
       \\for the KLOE-2 Collaboration
}

\institute{LNF-INFN, Via E. Fermi 40, 00044 Frascati (RM), Italy
          }

\abstract{The KLOE-2 experiment at the INFN Laboratori Nazionali di Frascati (LNF) is currently
taking data at the upgraded $e^{+}e^{-}$ DAFNE collider. Present Run II follows a~development
phase to assess the feasibility of a long term acquisition program, Run I,  which successfully
ended in July 2015 with 1 fb$^{-1}$ integrated luminosity collected in less than eight months.
KLOE-2 represents the continuation of the KLOE experiment with a new physics program.
The KLOE detector has undergone several upgrades including state-of-the-art cylindrical
GEM Inner Tracker, electron-positron taggers for the $\gamma\gamma$-physics studies
and new calorimeters around the interaction point.
In this article we briefly present the overview of the KLOE-2 experiment including the present
status and achievements together with the physics plans.
}
\maketitle
\section{Introduction}
\label{intro}
The KLOE-2 experiment operates at the upgraded DA$\Phi$NE $e^+e^-$ collider in Frascati aiming to
complete and extend the KLOE physics program with more than 5 fb$^{-1}$
of integrated luminosity collected in the next 2-3 years~\cite{AmelinoCamelia}.
For the first time the "`Crab-Waist"' interaction scheme developed in Frascati has been applied in the presence
of a high-field detector solenoid, where the transverse dimensions of the beams and their crossing angle
are tuned to maximize the machine luminosity~\cite{zobov}. KLOE-2 achieved a record performance in terms of 2 x 10$^{32}$
cm$^{-2}$s$^{-1}$ peak luminosity and 12 pb$^{-1}$ maximum daily integrated luminosity with this innovative
scheme of beam collisions, which will be employed in the upgrade of the B-factory currently under construction
at the KEK Laboratory in Japan, and is also considered a valid option in several future projects.\\
The main components of the KLOE-2 detector are the Drift Chamber, one of the biggest ever built, and
a lead-scintillating fiber Electromagnetic Calorimeter, one of the best ones for energy and timing performance
at low energies. The KLOE drift chamber provides three–dimensional tracking with perpendicular decay vertex
resolution of the order of 1~mm and transverse momentum accuracy better than 0.4$\%$~\cite{dc}. The electromagnetic
calorimeter consists of a 24-module barrel surrounding the drift chamber and two endcap detectors which together
cover in around 98$\%$ of the solid angle. It provides determination of energy and time with accuracies of
$\sigma_E/E = 5.7\%/\sqrt{E[\mathrm{GeV}]}$ and $\sigma(t) = 57\mathrm{ps}/
\sqrt{E[\mathrm{GeV}]} \oplus 100~\mathrm{ps}$, respectively~\cite{calo}. 	
 To improve vertex reconstruction capabilities near the interaction region KLOE-2 was equipped with an inner
tracker and it is the first high-energy experiment using the GEM technology with a cylindrical geometry,
a novel idea that was developed at LNF exploiting the kapton properties to build a transparent and compact
tracking system~\cite{it}. 
To study $\gamma\gamma$-physics the detector has been upgraded with two pairs of electron-positron taggers.
The Low Energy Tagger (LET) is a small calorimeter placed inside KLOE near the interaction piont, consisting
of LYSO crystals read out by silicon photomultipliers. This subdetector will serve to measure electrons and
positrons from $\gamma\gamma$ interaction with energy up to around 400~MeV~\cite{Babusci:2015osa}.
The second tagger is called High Energy Tagger (HET) which provides measurement of the displacement of
the scattered leptons with respect to the main orbit. Therefore, it was inserted inside the machine lattice
11~m away from the interaction point. This position sensitive detector consists of 30 small BC418 scintillators
3x3x5 mm$^3$, which provide a spatial resolution of 2~mm (corresponding to momentum resolution
of $\sim$1~MeV/c)~\cite{taggers}. In parallel to the ongoing data taking campaign we have completed several
analyses using the old KLOE data sample. Some of these results will be briefly described in the next section.
\section{Recent KLOE-2 results}
\label{results}
Apart from running the KLOE-2 experiment and activities devoted to commissioning of the new subdetectors
there are still several ongoing, or recently finished, physics analyses based on the KLOE data,
including search for dark forces at 1~GeV, tests of the CPT and Lorentz invariance in the neutral
kaon system and new precision measurements in hadronic physics at low energy.
\subsection{CPT tests in the neutral kaon system}
\label{kaons}
An unique feature of the $\Phi$-factory is the production of neutral kaon pairs in a pure quantum state,
thus at KLOE we can study quantum interference effects and tag pure monochromatic $K_S$ and $K_L$
beams. Due to the quantum entanglement the kaons double differential rate of decay into two final states
$f_1$ and $f_2$ takes the following form:
\begin{equation}
I(f_{1},f_{2}, \Delta t) \sim |\eta_{1}|^2 e^{-\Gamma_{L}\Delta t} + |\eta_{2}|^2 e^{-\Gamma_{S}\Delta t}
-2|\eta_{1}|^2|\eta_{2}|^2 e^{-\frac{(\Gamma_{L}+\Gamma_{S})}{2}\Delta t} \mathrm{cos}(\Delta m \Delta t + \phi_{2} -\phi_{1})
\label{interf}
\end{equation}
where $\Delta t$ denotes the proper decay times difference and $\eta_{1}$ and $\eta_{2}$
are the two decay amplitude ratios: $\eta_{i}$ = $|\eta_{i}|e^{i\phi_{i}} = \frac {\left\langle f_{i}|T|K_{L}\right\rangle}
{\left\langle f_{i}|T|K_{S}\right\rangle}$.
$\Gamma_{L}$ and $\Gamma_{S}$ denote the widths of $K_L$ and $K_S$ meson, respectively.
If both kaons decay to the same final state the interference
pattern in Eq.~\ref{interf} is very sensitive to any deviation from unity of the ratio
$\eta_{1}/\eta_{2}$ in the interference region of $\Delta t \approx$~0.
Such deviations may be present due to the CPT violation which in the framework of
SME manifests to lowest order only in the parameter $\delta_{K}$ exhibiting dependence
on the 4-momentum of the kaon~\cite{kost}:
\begin{equation}
\delta_{K} \approx i \mathrm{sin}\phi_{SW} e^{i\phi_{SW}}\gamma_{K}
(\Delta a_0 - \vec{\beta_K}\cdot \Delta\vec{a})/\Delta m~,
\label{deltak}
\end{equation}
where $\phi_{SW}$ is the superweak phase, $\gamma_{K}$ and $\vec{\beta_K}$ are the boost
factor and velocity of kaon and $\Delta m$ denotes the difference between $K_L$ and $K_S$
mass. The four $\Delta a_{\mu}$ parameters represent the differences of couplings to valence quark
and antiquark of the meson.
Since the Earth is rotating the CPT and Lorentz violating parameters are usually considered in
the fixed stars reference frame which makes $\delta_K$ dependent also on the sidereal time
$t_s$.
At KLOE a search for the Lorentz and CPT symmetries violation have been performed by measurements of
the interference pattern for both kaons decaying to $\pi^+\pi^-$ final states using 
a sample of 1.7 fb$^{-1}$ of integrated luminosity~\cite{rumcajs}.
\begin{figure}
  \begin{center}
	\includegraphics[width=0.7\textwidth]{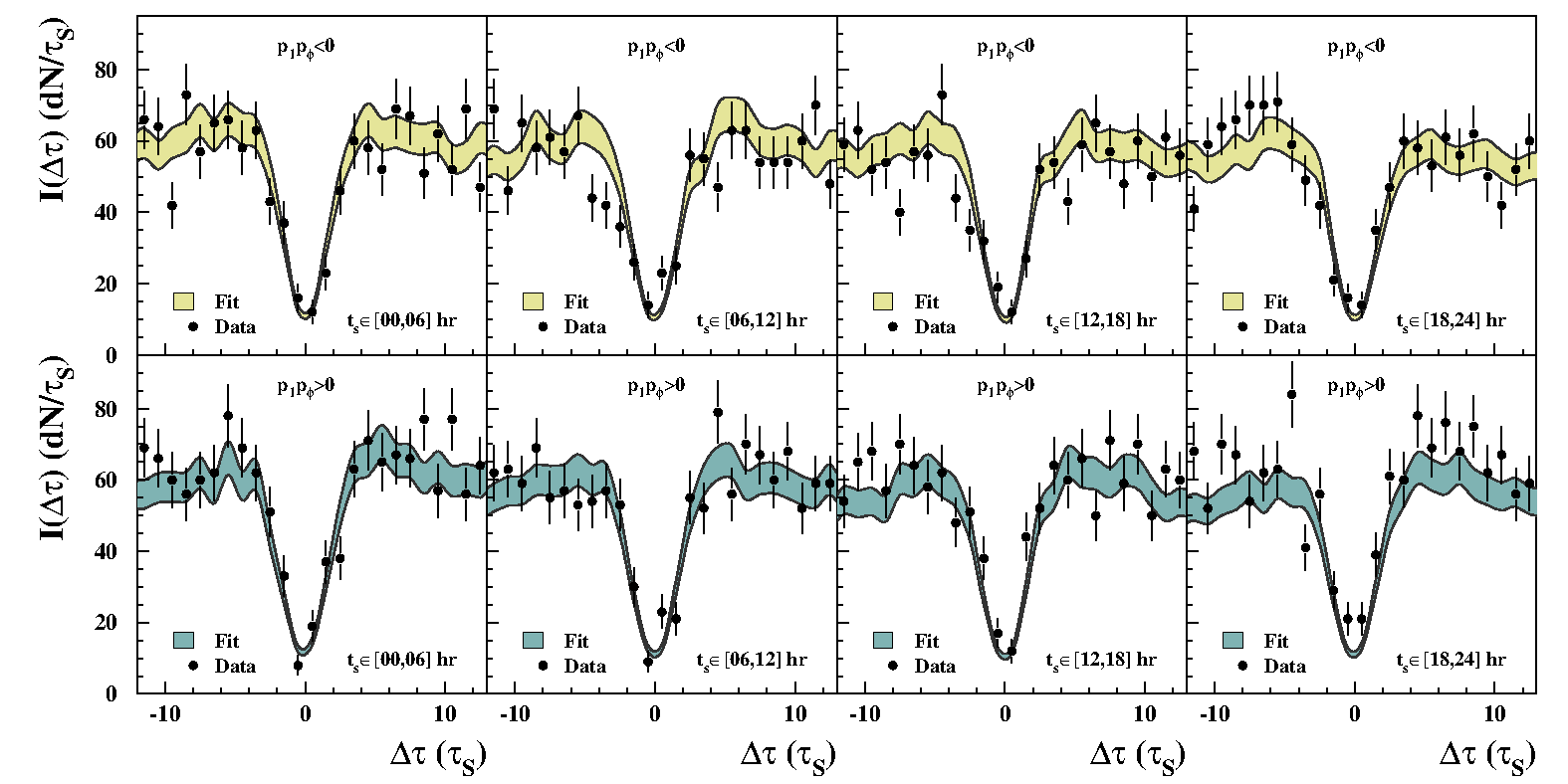}
  \end{center}
  \caption{Distributions of the double differential decay rates into $\pi^+\pi^-,\pi^+\pi^-$ final states
	measured with the KLOE detector. The top panel contains distributions for four bins of sidereal time
	when kaon emitted in the forward hemisphere ($\mathrm{cos}\theta > 0$) in the KLOE reference
frame was moving along with the $\phi$ momentum (i.e. $\vec{p}_{1(2)}\cdot \vec{p}_{\phi}>0$).
The lower panel shows distributions for a subset of data where kaon emitted in the forward hemisphere
was moving in the opposite direction (i.e. $\vec{p}_{1(2)}\cdot \vec{p}_{\phi} < 0$). Figure taken
from~\cite{rumcajs}.} 
  \label{figfit}
\end{figure}
The data sample was divided into two subsets: one containing events where kaon emitted in the forward
hemisphere in the KLOE reference frame was moving along with the $\phi$ momentum and the second one
with events where kaon emitted in the same hemisphere was moving in the opposite direction. Moreover,
the data were divided into four bins of sidereal time. For every subset of the data distribution of
the decay intensity in function of $\Delta t$, expressed in units of the $K_S$ lifetime, was determined.
The resulting eight distributions are presented in~Fig.~\ref{figfit}. The distributions were fitted
simultaneously with the interference pattern allowing possible modulation effects induced by the CPT
violation. The results of the fit are indicated in Fig.~\ref{figfit} together with uncertainties
by colored bands. They do not reveal any CPT or Lorentz symmetries violation and constitute the first
estimation of $\Delta a_{\mu}$ parameters in the kaon sector with sensitivity of about 10$^{-18}$~GeV~\cite{rumcajs}.
\subsection{Search for dark forces}
\label{darkB}
In the framework of the Dark Matter searches it was postulated, that there may exist a secluded gauge sector
mediated by a new vector gauge boson $U$, called also the dark photon~\cite{dark1}. The dark photon could couple
with ordinary photon via mixing term and subsequently decay into Standard Model particles giving opportunity to
be observed in high-energy experiments. Direct searches of these dark matter mediators are performed with several
experiments including KLOE-2.  We have explored three different processes: Dalitz decays of the $\phi$ meson:
$\phi \rightarrow \eta U,~U \rightarrow e^+e^-$ with $\eta \rightarrow 3\pi$~\cite{kloed1,kloed2},
$e^+e^- \rightarrow U\gamma$ with $U$ decaying to lepton or pion pairs~\cite{kloed3,kloed4,kloed5}
and the dark Higgsstrahlung process $e^+e^-\rightarrow U h_0,~U \rightarrow \mu^+\mu^-$
with $h_0$ invisible~\cite{kloed6}.
\begin{figure}
\begin{center}
\sidecaption
\includegraphics[width=0.5\textwidth]{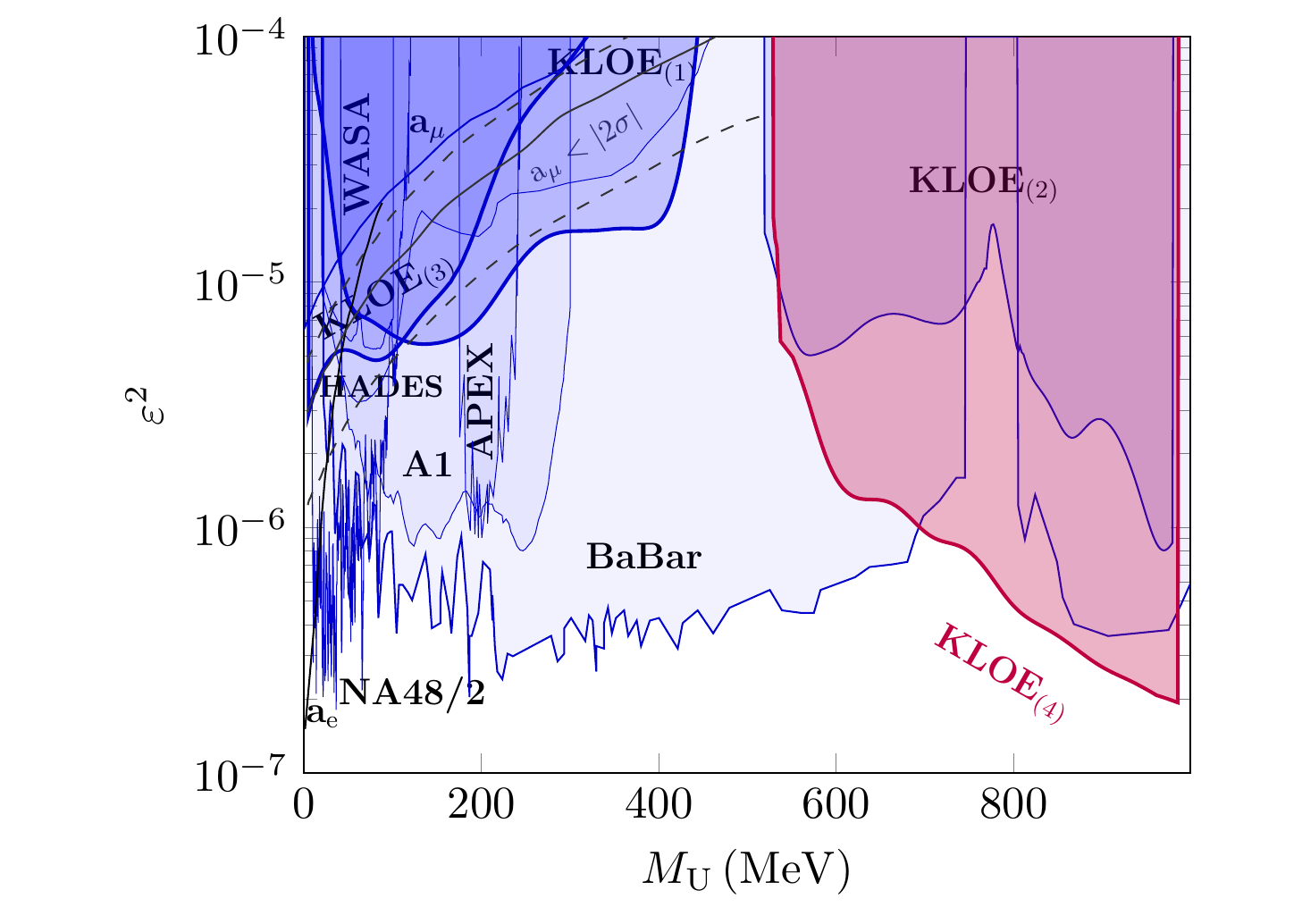}
\caption{
90\% CL exclusion plot for the $U$ boson coupling $\varepsilon^2$ as a function of the $\mathrm{U}$-boson
mass. The limits obtained by KLOE-2 are denoted as: KLOE$_{(1)}$ ($\phi$ Dalitz dacay studies~\cite{kloed1,kloed2}),
KLOE$_{(2)}$ ($e^+e^- \to e^+e^-\gamma$~\cite{kloed3}), KLOE$_{(3)}$ ($e^+e^- \to \mu^+\mu^-\gamma$~\cite{kloed4})
and KLOE$_{(4)}$ ($e^+e^- \to \pi^+\pi^-\gamma$~\cite{kloed5}).
The limits from  the A1~\cite{Mami1}, APEX~\cite{Apex}, the WASA~\cite{WASA}, HADES~\cite{HADES}, BaBar~\cite{BaBar}
and NA48/2~\cite{NA48/2} experiments are also shown. The gray dashed lines shows $U$ boson parameters that could explain
the observed muon anomalous magnetic moment~\cite{a_mu}. Figure adapted from~\cite{kloed6}.
}
\label{UL}
\end{center}
\end{figure} 	
Search for the dark photon in the $\phi$ Dalitz dacays were performed using an integrated luminosity
of 1.7~fb$^{-1}$ and 1.5~fb$^{-1}$
for $\eta \to 3\pi^0$ and $\eta \to \pi^+\pi^-\pi^0$, respectively. 
The irreducible background from the $\phi \to \eta \gamma^{*} \to \eta e^+e^-$ decay was directly extracted from the
data by a fit to the di-lepton invariant mass distribution parameterized according to the Vector Meson Dominance model.
No resonant signal was observed in the $e^+e^-$ invariant mass distributions for both $\eta$ decays. As a result we have estimated
upper limit on the kinetic mixing parameter as a function of the $U$ boson mass which is shown in Fig.~\ref{UL} as the KLOE$_{(1)}$
region.\\
We have searched for the $e^+e^- \rightarrow U\gamma$ analysing three possible $U$ decays into $e^+e^-$, $\mu^+\mu^-$ and $\pi^+\pi^-$.
Again the signature of the dark photon is the narrow structure of the lepton-antilepton or $\pi^+\pi^-$ invariant mass distribution.
In case of the $\mu^+\mu^-\gamma$ and $\pi^+\pi^-\gamma$ we were looking for both charged particles emitted with large polar angle and
an undetected photon at a small angle which resulted in a strong suppression of the resonant $\phi \to \pi^+\pi^-\pi^0$ and FSR
background. The residual background components were simulated and fitted to the invariant mass spectra. The measured distributions
are in perfect agreement with the simulations done with the PHOKHARA generator~\cite{phok} and does not show any peak structure.
The exclusion limits estimated based on these studies are shown in Fig.~\ref{UL} as KLOE$_{(2)}$, KLOE$_{(3)}$ and KLOE$_{(4)}$
regions.
A natural consequence of massive dark photon would be the existence of associated Higgs-like particle $h'$, so-called dark Higgs,
coupling to U with coupling constant $\alpha_D$. Thus, the U boson could manifest itself in the $e^+e^-$ collisions through
the Higgsstrahlung process: $e^+e^- \to Uh'$ where $U$ decays to the $\mu^+\mu^-$ pair. If the mass of $h'$ exeeds mass of
the $U$ boson the transition $h' \to U$ is allowed and one can search for the Standard Model decay products of both $U$ bosons
produced (e.g. two di-lepton or di-pion pairs)~\cite{belle,babar}. The KLOE-2 analysis was done assuming $m_{h'} < m_U$, therefore
we were looking for the missing mass in the final state of $\mu^+\mu^-$~\cite{kloed6}. The expected signal would show up as a sharp
enhancement in the  two dimensional distribution of $\mu^+\mu^-$ missing mass versus the $\mu^+\mu^-$ invariant mass.
In order to better understand backgrounds, two data samples were analysed: sample gathered at the $\phi$ mass peak (1.65~$fb^{-1}$)
and the off-peak sample at $\sqrt{s}~ =$~1000~MeV (0.2~$fb^{-1}$), where backgrounds are strongly reduced. 
The 90$\%$ confidence level limits of $\alpha_D \cdot \varepsilon^2$ obtained by KLOE-2 for combined on- and off-peak data samples
are at the level of $10^{-9}-10^{-8}$ depending on the masses of $U$ and $h'$~\cite{kloed6}.
\subsection{New results in hadronic physics}
\label{hadrony}
Recently KLOE-2 published new, independent result on the $\eta \to \pi^+\pi^-\pi^0$ Dalitz
plot measurement with the highest statistics in the world ($\sim 4.48\cdot 10^6$ events)~\cite{Li}.
This isospin-breaking decay is of particular interest because it provides information about the light-quark
mass difference. Moreover, analysis of the $\eta \to \pi^+\pi^-\pi^0$ Dalitz plot gives possibility
to test the charge conjugation symmetry.
At KLOE the $\eta$ meson is produced in the radiative $\phi$ decay together with a mono-energetic photon with
energy E~$\sim$~363~MeV. The candidate events are searched for in a preselected sample with at least three prompt
photons and two tracks with opposite curvature reconstructed. The main background originate from the $e^+e^- \to \omega\pi^0$
reaction and the Bhabha scattering.
The resulting $\eta \to \pi^+\pi^-\pi^0$ Dalitz plot was bin-by-bin background subtracted and fitted with
the decay squared amplitude $|A|^2$ parametrized with a polynomial expansion:
\begin{eqnarray}\label{eq:DPamplitude}
|A(X,Y)|^2&\simeq& N(1+aY+bY^2+cX+dX^2+eXY + fY^3+ gX^2Y + \ldots),
\end{eqnarray}
where $X$ and $Y$ are the normalized Dalitz plot variables expressed by the kinetic energies of all the particles
in the final state~\cite{Li}. We have obtained the most precise estimations of the Dalitz parameters, for
the first time including also the $g$ parameter. The final values of the charge asymmetries are all consistent
with zero and were obtained with the best sensitivity in the world~\cite{Li}.\\
Precise determination of the transition form factors (TFF) of mesons is crucial in many fields
of particle physics, such as determination of the light-by-light contribution to the anomalous magnetic
moment of the muon or understanding the low-energy structure of hadrons.
KLOE-2 has obtained new results on TFFs and branching fractions for the $\phi \to \eta e^+e^- \to 3\pi^0 e^+e^-$
and $\phi \to \pi^0 e^+e^-$ decays~\cite{etaee,pi0ee}. 
\begin{figure}
\begin{center}
\includegraphics[width=0.47\textwidth]{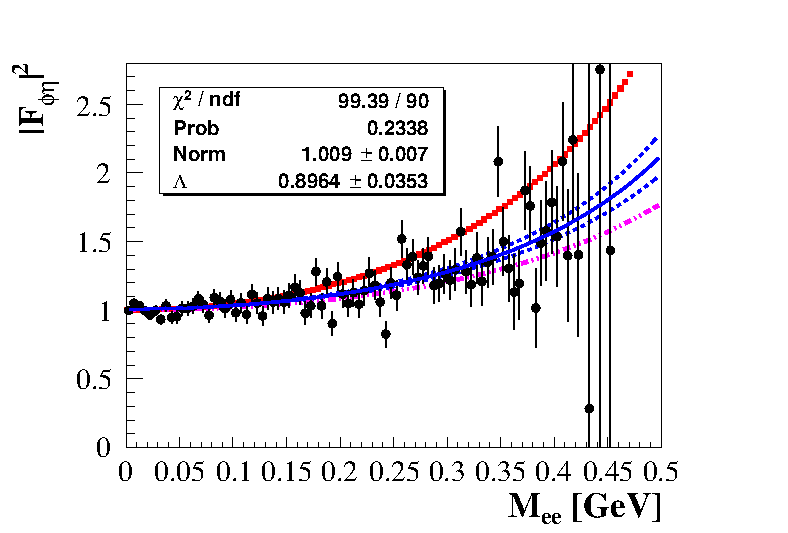}
\includegraphics[width=0.47\textwidth]{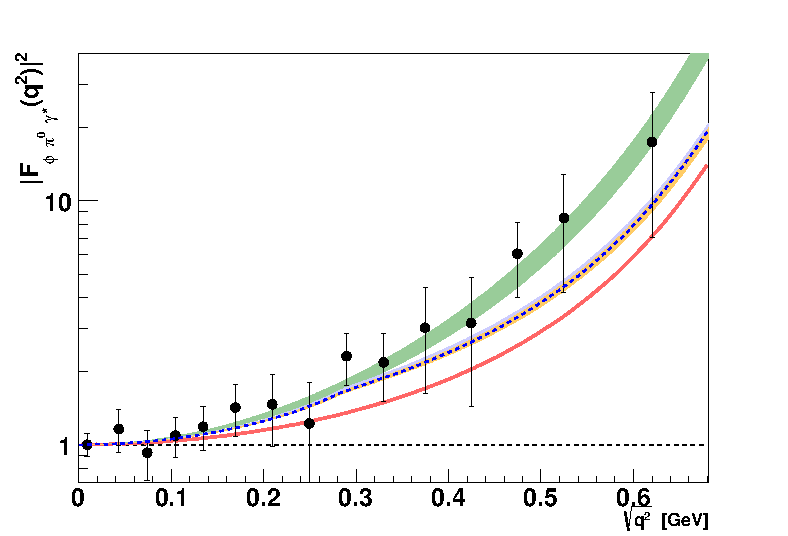}
\caption{Left: $F_{\phi \, \eta}$ distribution as a function of the $e^+e-$ invariant mass.
The blue curve shows the results of fit of the VMD model parametrization to the data.
VMD theoretical expectations are superimposed in pink dashed curve while the $F_{\phi \, \eta}$
dependence obtained from the Terschluesen/Leupold model is reported as red points.
Right: Measured $F_{\phi \, \pi^0}$ in function of modulus of the 4-momentum transfer
$\sqrt{q^2}$ (black points)
and the theoretical predictions based on: the dispersive analysis of Ref.~\cite{disp1}
(orange and cyan bands) and Ref.~\cite{disp2} (blue dashed line), the chiral theory approach~\cite{chiral}
(green band), and the one-pole VMD model~\cite{disp2} (solid red line).}
\label{ffs}
\end{center}
\end{figure}
In the case of $\phi \to \eta e^+e^-$ decay the Form Factor $F_{\phi \,\eta}$ was extracted
from the $e^+e^-$invariant mass spectrum by dividing the measured invariant mass spectrum
by the spectrum of reconstructed MC signal events generated with a constant $F_{\phi \, \eta}$,
after all the analysis cuts. The resulting $F_{\phi \, \eta}$ distribution has been fitted
as a function of the invariant mass with the Vector Meson Dominance model (VMD) parametrization~\cite{vmd}.
The result of the fit is shown in the left panel of Fig.~\ref{ffs} together with theoretical
predictions from the VMD (pink dashed curve) and Terschluesen/Leupold models~\cite{leupold}.
The value of the TFF slope obtained with the fit amounts to $b_{\phi\eta} = 1.28 \pm 0.10^{+0.09}_{-0.08}$~GeV$^{-2}$~\cite{etaee} 
and is in agreement with the VMD predictions~\cite{vmd}.
An analogus analysis of the $\phi \to \pi^0 e^+e^-$ decay led to the determination of the modulus square
of the $F_{\phi \, \pi^0}$ Transition Form Factor in a function of the 4-momentum modulus $\sqrt{q^2}$ below 700~MeV~\cite{pi0ee}
shown in the right panel of Fig.~\ref{ffs} together with theoretical expectations. As one can see the chiral theory
approach~\cite{chiral} is favoured by the data. The TFF slope parameter extracted from this dependence amounts to
$b_{\phi\pi^{0}} = 2.02 \pm 0.11$~GeV$^{-2}$.

\section{Summary}
\label{summary}
The KLOE-2 experiment constitute the continuation of KLOE with a new physics program mainly focused on
the study of $K_s$, $\eta$ and $\eta'$ decays as well as on kaon interferometry, test of discrete symmetries
and search for physics beyond the Standard Model~\cite{AmelinoCamelia}. The new data taking campaign aiming to collect more than
5~fb$^{-1}$ integrated luminosity in the next 2-3 years, will allow to perform CPT symmetry and quantum
coherence tests using entangled neutral kaons with an unprecedented precision, high precision studies
of $\gamma\gamma$-physics processes like $e^{+}e^{-}\to e^{+}e^{-} \pi^0 (\gamma\gamma \to \pi^0)$,
and the search for signals of a hidden dark-matter sector, among the fields to be addressed.
\\
\begin{acknowledgement}
This work was supported in part by the EU Integrated Infrastructure Initiative Hadron Physics Project under
contract number RII3-CT- 2004-506078; by the European Commission under the 7th Framework Programme through
the `Research Infrastructures' action of the `Capacities' Programme, Call: FP7-INFRASTRUCTURES-2008-1,
Grant Agreement No. 227431; by the Polish National Science Centre through the Grants No.\
2011/03/N/ST2/02652,
2013/08/M/ST2/00323,
2013/11/B/ST2/04245,
2014/14/E/ST2/00262,
2014/12/S/ST2/00459.
\end{acknowledgement}
%
%
%

\end{document}